\documentclass[sigconf]{acmart}
\usepackage{xspace}
\usepackage{url}

\newcommand{\modelname}{RecNextEval\xspace}

\newcommand{\eg}{\emph{e.g.,}\xspace}

\AtBeginDocument{%
  }

\copyrightyear{2026}
\acmYear{2026}
\setcopyright{cc}
\setcctype{by}
\acmConference[SIGIR '26]{Proceedings of the 49th International ACM SIGIR Conference on Research and Development in Information Retrieval}{July 20--24, 2026}{Melbourne, VIC, Australia}
\acmBooktitle{Proceedings of the 49th International ACM SIGIR Conference on Research and Development in Information Retrieval (SIGIR '26), July 20--24, 2026, Melbourne, VIC, Australia}
\acmDOI{10.1145/3805712.3808369}
\acmISBN{979-8-4007-2599-9/2026/07}

\begin{document}

\title{RecNextEval: A Reference Implementation for Temporal Next-Batch Recommendation Evaluation}

\author{Tze-Kean Ng}
\email{tng042@e.ntu.edu.sg}
\orcid{0009-0007-6894-199X}
\affiliation{%
  \institution{Nanyang Technological University}
  \city{Singapore}
  \country{Singapore}
}

\author{Joshua Teng-Khing Khoo}
\email{jkhoo013@e.ntu.edu.sg}
\orcid{0009-0007-6656-1336}
\affiliation{%
  \institution{Nanyang Technological University}
  \city{Singapore}
  \country{Singapore}
}

\author{Aixin Sun}
\authornote{Corresponding author.} 
\email{axsun@ntu.edu.sg}
\orcid{0000-0003-0764-4258}
\affiliation{%
  \institution{Nanyang Technological University}
  \city{Singapore}
  \country{Singapore}
}

\begin{abstract}
A good number of toolkits have been developed in Recommender Systems (RecSys) research to promote fair evaluation and reproducibility. However, recent critical examinations of RecSys evaluation protocols have raised concerns regarding the validity of existing evaluation pipelines. In this demonstration, we present \modelname, a reference implementation of an evaluation framework specifically designed for next-batch recommendation. \modelname utilizes a time-window data split to ensure models are evaluated along a global timeline, effectively minimizing data leakage. Our implementation highlights the inherent complexities of RecSys evaluation and encourages a shift toward model development that more accurately simulates production environments. The \modelname library and its accompanying GUI interface are open-source and publicly accessible.\footnote{ \modelname library: \url{https://github.com/hiiamtzekean/recnexteval} \newline
 User interface:   \url{https://github.com/hiiamtzekean/recnexteval-studio}.} 
\end{abstract}



\maketitle

\section{Introduction}
\label{sec:intro}
Recommendation systems have evolved from simple convenience tools into the invisible backbone of global consumption. The field is advancing rapidly, driven by a shift from basic collaborative filtering methods to sophisticated deep learning architectures capable of decoding human intent in real time. While much research focuses on algorithmic advancement, academic work has remained largely constrained to widely used yet increasingly outdated datasets and, more critically, to offline evaluation settings~\cite{EvaluationLandscape24}.

Several large-scale evaluations have been conducted~\cite{McElfreshKV0W22}, and numerous toolkits~\cite{sun2022daisyrec,AnelliBFMMPDN21,ZhuDSMLCXZ22,Ekstrand20,Argyriou20,recbole[1.2.1],li2024rechorus2,Chia22,Michiels22} have been developed to support RecSys research.\footnote{\url{https://github.com/ACMRecSys/recsys-evaluation-frameworks}} However, an examination of these toolkits through their documentation and codebases reveals that existing implementations primarily emphasize predictive models and increasingly comprehensive evaluation metrics, while often overlooking temporal aspects in their design. In particular, many implementations assume that traditional data splitting (\eg random splits and user-based leave-last-one-out) and evaluation protocols are sufficient, thereby failing to account for the dynamic nature of user behavior over time.

More recently, research has increasingly recognized the importance of preserving a \textit{global timeline} during evaluation and has adopted fixed time-point splits to mitigate data leakage~\cite{Yambda5b25,Klenitskiy26}.\footnote{Data leakage occurs when future interactions are inadvertently made accessible to the RecSys model  during training, a scenario that would not be possible in  production environments.} A few toolkits like RecPack~\cite{Michiels22} and DaisyRec 2.0~\cite{sun2022daisyrec} also support data split by fixed timestamp. While this approach addresses leakage concerns, it introduces two additional limitations. First, evaluation results may depend heavily on the chosen single split point. Second, user preferences can evolve over time due to trends, seasonality, or external factors—a phenomenon known as concept drift~\cite{ConceptDrift14}. Fixed time-point evaluations fail to adequately reflect a model’s ability to capture and adapt to such temporal dynamics.

We introduce \modelname, a reference implementation for temporal next-batch recommendation evaluation. \modelname does not provide implementations of recommendation algorithms or models; instead, it serves as an evaluation platform that constrains data release to evaluated  RecSys models in order to enforce a global timeline. Through APIs, \modelname closely simulates a production environment where new data arrives continuously and the RecSys model must be iteratively updated and tested against incoming requests. RecNextEval fundamentally differs from existing toolkits through its sliding window paradigm, which produces \textit{multiple evaluation rounds} across a global timeline. 

For datasets, to be evaluated on \modelname, all user–item interactions are assumed to be associated with timestamps. RecSys models evaluated on the platform must implement a set of interface APIs to communicate with \modelname. Through these interfaces, a model first receives an initial collection of user–item interactions to construct a base model. The evaluation then proceeds sequentially over a series of time windows, similar to test-then-train experimental protocol over data streams~\cite{EvaluationStream15}. In each time window, the model is first evaluated on its ability to generate recommendation predictions for users specified by \modelname. After prediction, the model receives the corresponding ground truth, along with the remaining user–item interactions occurring within that window. The model is expected to \textit{incrementally learn} from this newly available data and prepare for the next round of prediction in the subsequent time window, until the end of the data stream. Finally, \modelname reports evaluation results across all time windows and outputs the overall performance using both macro- and micro-averaged metrics including Hit Rate and NDCG.

\modelname simulates a data stream in an online setting and provides an opportunity for RecSys researchers to better understand the practical requirements of recommender system models from the perspectives of cold-start, incremental learning, and related challenges including evaluation along timeline. 
\section{Next-Batch Recommendation Evaluation}
\label{sec:design}

\begin{figure}
 \center
   \includegraphics[clip, trim=0.3cm 8.2cm 6.7cm 3.2cm, width=\linewidth]{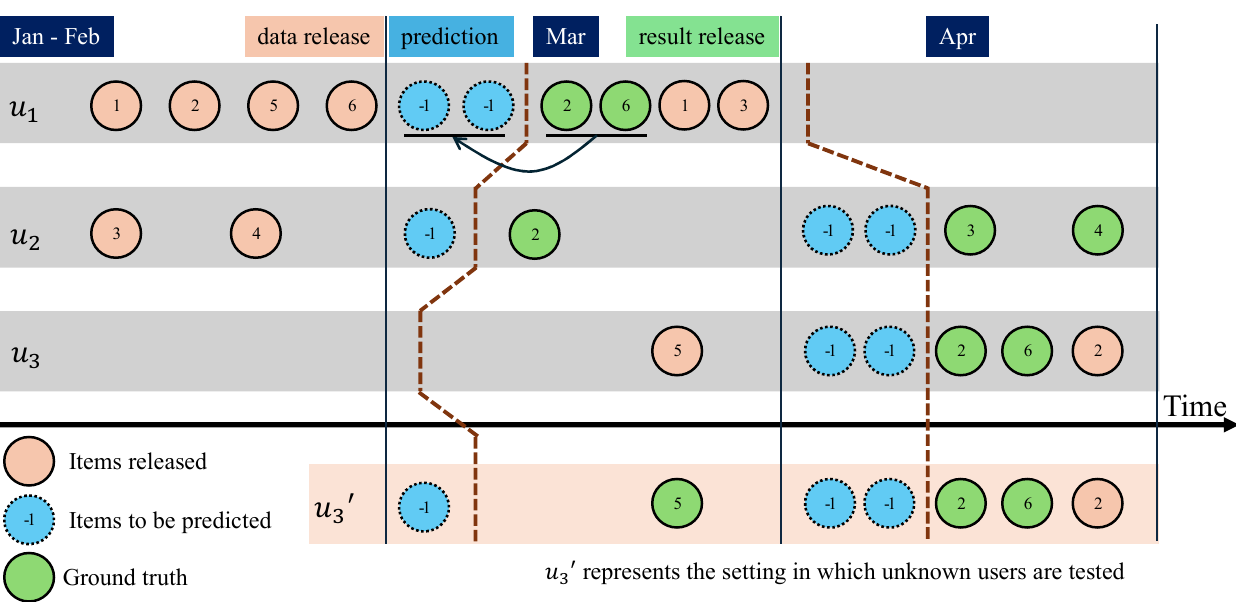}
  \caption{An illustration of the data release, prediction, and result release phases using three example users. Items interacted with by each user are plotted along the timeline. Data from Jan to Feb are released to build the base model, and within each time window (e.g., Mar), the model is requested to generate predictions before receiving the ground-truth items along with the remaining user interactions.}
  \Description{Illustration of data release, prediction, and result phases for three users, showing timelines and item interactions.}
  \label{fig:slideSetting}
\end{figure}

The design of the next-batch recommendation evaluation is primarily based on the Timeline scheme proposed in~\cite{FreshSun23}. In this framework, all user-item interactions are associated with a specific timestamp and organized along a global timeline. We use Figure~\ref{fig:slideSetting} to illustrate the rationale behind our evaluation design, the specific challenges of temporal data, and our proposed solutions.

As shown in Figure~\ref{fig:slideSetting}, interactions for three example users ($u_1$, $u_2$, and $u_3$) are plotted against a global timeline partitioned into distinct windows (e.g., "Jan-Feb", "Mar", and "Apr").

Initially, interaction data from the "Jan-Feb" window are released to the recommender model for training, to build an initial model. All subsequent data are withheld by \modelname to prevent temporal data leakage. Note that, at this stage, user $u_3$ has no recorded interactions; thus, $u_3$ is an unknown user to the model, or "cold-start" user to the model in the future.

\subsection{Prediction and Evaluation}
\label{sec:pred_and_eval}

For the data occurring in the "Mar" window, \modelname operates in two distinct phases:

\vskip 1ex\noindent{\textbf{Prediction Phase}}: The framework requests a model being evaluated to make recommendations for a set of user IDs. The maximum number of predictions per user is configurable, which is 2 in Figure~\ref{fig:slideSetting} as an illustration. Specifically, $u_1$ is assigned two predictions, indicated by the blue circles labeled "-1". The number of requests adapts to actual user activity. While $u_1$ receives two requests, $u_2$ only has one interaction during this window, so only one prediction is requested. Each request is accompanied by the exact timestamp of the actual interaction to maintain temporal integrity. In this  phase, the model only receives these user-based prediction requests.

User $u_3$ presents a distinct scenario compared to $u_1$ and $u_2$. Although $u_3$ interacts with an item during the March window, no prior training data exists for this user. To handle such cases, \modelname offers two configurable settings: \textit{enabling} or \textit{disabling} the evaluation of \textit{unknown users}. In the scenario illustrated in Figure~\ref{fig:slideSetting}, we have opted to disable unknown user evaluation. Consequently, no prediction requests are sent to the recommender model for $u_3$ during the March prediction phase.

The predictions made by the RecSys model for the requested users are then submitted to \modelname for evaluation metric computation, both for this time window and the entire evaluation.

\vskip 1ex\noindent{\textbf{Result Release Phase}}. The second operation phase within the March window is the Result Release phase, where two types of data are made available to the model.

\paragraph{Ground Truth} The framework releases the actual items corresponding to previous prediction requests. For instance, as shown in Figure \ref{fig:slideSetting}, items 2 and 6 are revealed as the ground truth for $u_1$'s two prediction requests which was originally marked as "-1".

\paragraph{Remaining Interactions} Since a user may have more interactions within a time window than there were prediction requests, \modelname also releases all remaining interaction data. The model is expected to \textit{incrementally learn} from this "new batch" of available data (including both ground truth and newly released interactions) to prepare for the subsequent evaluation window. In the case of $u_2$, who only had a single interaction in March, only the ground truth is released. For $u_3$, the single interaction made in March is now released to the model. Consequently, from this point forward, $u_3$ is no longer an unknown user, and their interactions in April will be evaluated accordingly. 

This two-phase operation repeats for the April window. Notably, as $u_1$ has no interactions during April, no prediction requests are generated for this user, illustrating how the framework dynamically adapts to user activity over the global timeline.

Recall that our discussion so far assumed a setting in which unknown-user evaluation is disabled. User $u_3'$ (plotted below the timeline in Figure~\ref{fig:slideSetting}) represents the alternative configuration, where unknown-user evaluation is enabled. The model is asked to generate predictions for $u_3'$ during the March prediction phase, even though no prior interactions are available to the model at that time.

As discussed for unknown users, \textit{the same principle applies to unknown items}. Unknown items may be excluded during prediction so that evaluation is performed only on exposed items. \modelname provides configuration options to ignore unknown users and items, supporting multiple evaluation modes depending on the developer’s use case.

\subsection{Evaluation Metric Computation}

\modelname evaluates the accuracy of recommendations produced by the models at different levels. Note that accuracy here refers to the quality of the recommendations generated after the prediction phase and can be reflected by multiple metrics, such as Precision@$k$, Recall@$k$, Hit Rate@$k$, NDCG, and others. In the following discussion, we use Hit Rate@$k$ (HR) as an example metric.

\begin{itemize}
    \item \textbf{User-level HR within a window:} 
    A user (e.g., $u_1$ in Figure~\ref{fig:slideSetting}) may have multiple prediction requests. We first compute the Hit Rate for user $u$ within window $w$, denoted as $HR_u^w$.

    \item \textbf{Window-level HR:} 
    Considering all users evaluated within a given test window, the window-level Hit Rate is computed as $    HR_w = \frac{1}{|U_w|} \sum_{u \in U_w} HR_u^w,$
    where $U_w$ denotes the set of users evaluated in window $w$.

    \item \textbf{Macro HR:} 
    If we consider each window contributes equally to the overall evaluation across all windows, the macro-level Hit Rate is defined as the average over all test windows: $HR_M = \frac{1}{|W|} \sum_{w \in W} HR_w$, where $W$ denotes the set of all test windows.

    \item \textbf{Micro HR:} 
    If each evaluated user in a window contributes equally to the overall evaluation across all windows, the micro-level Hit Rate is defined as $HR_m = \frac{\sum_{w \in W} \sum_{u \in U_w} HR_u^w}{\sum_{w \in W} |U_w|}$.
\end{itemize}

\section{\modelname Implementation}
\label{sec:implementation}

Following the evaluation design, a RecSys model evaluated on \modelname under a streaming setting must implement a few key interface functions:
\textsf{register\_model()}, \textsf{get\_training\_data()}, \textsf{fit()}, \textsf{get\_unlabeled\_data()}, \textsf{predict()}, and \textsf{submit\_prediction()}.

\begin{figure}
 \center
  \includegraphics[clip, trim=0.8cm 6.5cm 15.5cm 3.2cm,  width=0.9\linewidth]
  {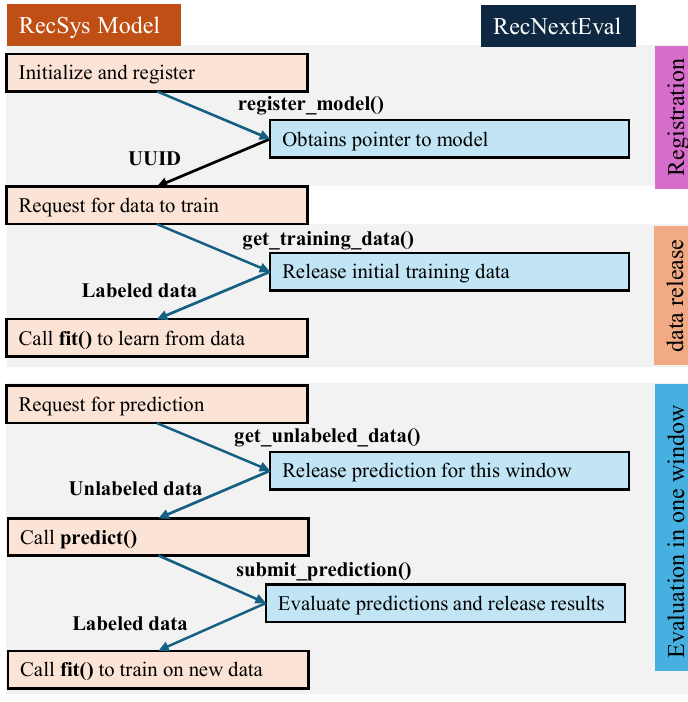}
  \caption{An illustration of the key APIs implemented by a RecSys model when evaluated on the \modelname platform. \modelname performs the evaluation in each time window, and report the aggregated results upon completion.}
  \label{fig:slideCommunication}
  \Description{An illustration of the key APIs implemented by a RecSys model when evaluated on the \modelname platform}
\end{figure}

As shown in Figure~\ref{fig:slideCommunication}, the model initializes the evaluation process by sending a registration request via \textsf{register\_model()}. \modelname then returns a UUID, stores the model metadata, and begins tracking the model’s evaluation status. After registration, the model receives an initial batch of training data during the data release phase, which is used to learn an initial or base model before any incremental updates occur.

Once learned, the RecSys model proceeds to the prediction phase for the first test window by requesting users and associated interaction timestamps through the \textsf{get\_unlabeled\_data()} interface. \modelname verifies the model’s current status and releases the set of users along with the timestamps for which recommendations must be generated, according to the specified evaluation configuration (e.g., handling of unknown users and unknown items). The model then makes the predictions by calling \textsf{predict()}.

After the model submits its predictions with \textsf{submit\_prediction()}, \modelname evaluates the submitted results and releases the corresponding ground truth, as well as the remaining user interactions within the window, as shown in Figure~\ref{fig:slideSetting}. These newly revealed interactions are then used by the model to incrementally update its parameters via the \textsf{fit()} function. The same sets of interface calls will be made by the model till all evaluation windows are completed: \textsf{get\_unlabeled\_data()}, \textsf{predict()},  \textsf{submit\_prediction()}, and  \textsf{fit()}.

\subsection{Python Package and Web UI}

\modelname is distributed as a Python package with ten modules, including core components: datasets, settings, evaluators, and metrics. Where necessary for greater modularity, sub-modules also used such as in settings and evaluators that provide greater code clarity. Our design employs OOP for separation of concerns, optimizations like caching and SciPy sparse matrices to reduce memory overhead and computation time, and design patterns (Strategy, Builder, Iterator) for extensibility and facilitation of contributions from the community. The code comes with dynamic documentation to further aid developers.

RecNextEval-Studio provides a web-based interface for RecNextEval, allowing developers and researchers to abstract away complex API interactions through an intuitive user interface for streamlined model evaluation. Deployed as a Docker-orchestrated stack, it integrates PostgreSQL for data persistence, a Python FastAPI backend, and a Vite-powered React frontend, enabling end-to-end evaluation of streaming recommender systems. In addition to hosted deployments, developers can also run the application on-premise by cloning the repository and launching the stack via Docker.

The web frontend follows conventional web development practices, using JavaScript and Tailwind CSS for styling. The implementation  offers authentication-aware navigation and user interfaces that support the core functionalities currently provided by \modelname, with an emphasis on usability and intuitive workflows.

The web backend exposes versioned REST APIs and supports full CRUD operations for communication with the frontend service. FastAPI manages persistence of user inputs in PostgreSQL and executes \modelname workflows by spawning background threads to handle user-submitted streaming evaluation jobs. This architecture supports  concurrency, allowing users to submit multiple requests while background tasks—managed by \textsf{Starlette}—execute asynchronously, without blocking the user interface or requiring users to wait synchronously for evaluation runs to complete.

\subsection{Reproducibility}

To ensure reproducibility, \modelname provides full package versioning on \texttt{GitHub}, along with dependency version control using \texttt{uv}, ensuring that usage is independent of the underlying machine environment. Furthermore, all implementations rely exclusively on open-source tools, allowing developers to fully reproduce experimental results and outcomes.

Versioned releases of \modelname are published on PyPI and can be installed using compatible Python package managers. Additional guidance and documentation are available in the GitHub repository.

To set up the system, users can follow the instructions in the \texttt{README.md} file in the repository. Cloning the repository and running the provided \texttt{make} command will start the required services for the web application. Alternatively, developers who only wish to use the programmatic API can install the package directly via \texttt{uv add recnexteval}, which installs all required dependencies.
\section{Demonstration}
\label{sec:interface}

\begin{figure}
 \center
  \includegraphics[width=0.85\linewidth]{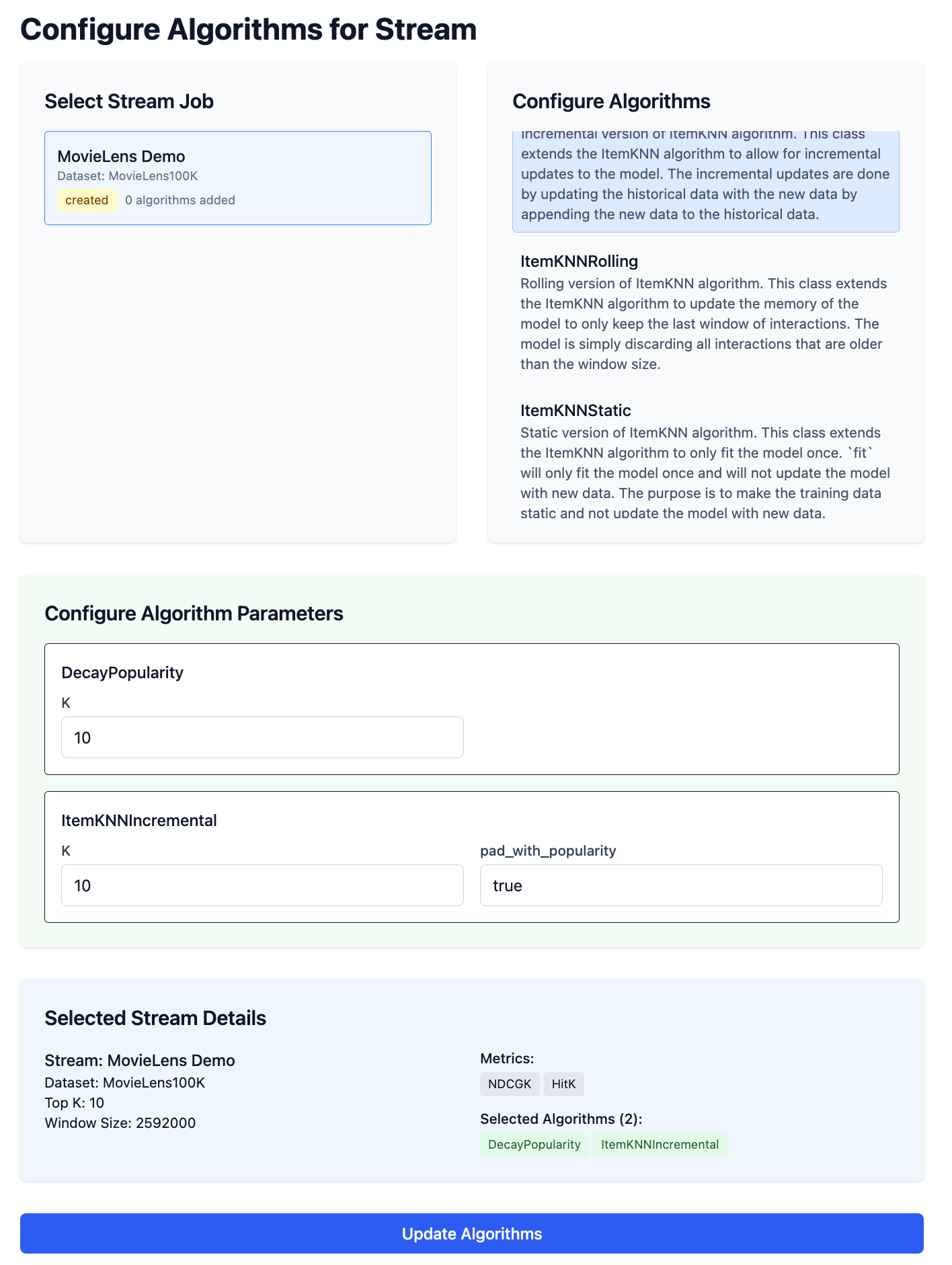}
  \caption{Configuring sample models for evaluation.}
  \Description{Configuring sample models for evaluation.}
  \label{fig:GUI-config}
\end{figure}

\begin{figure}
 \center
  \includegraphics[width=0.85\linewidth]{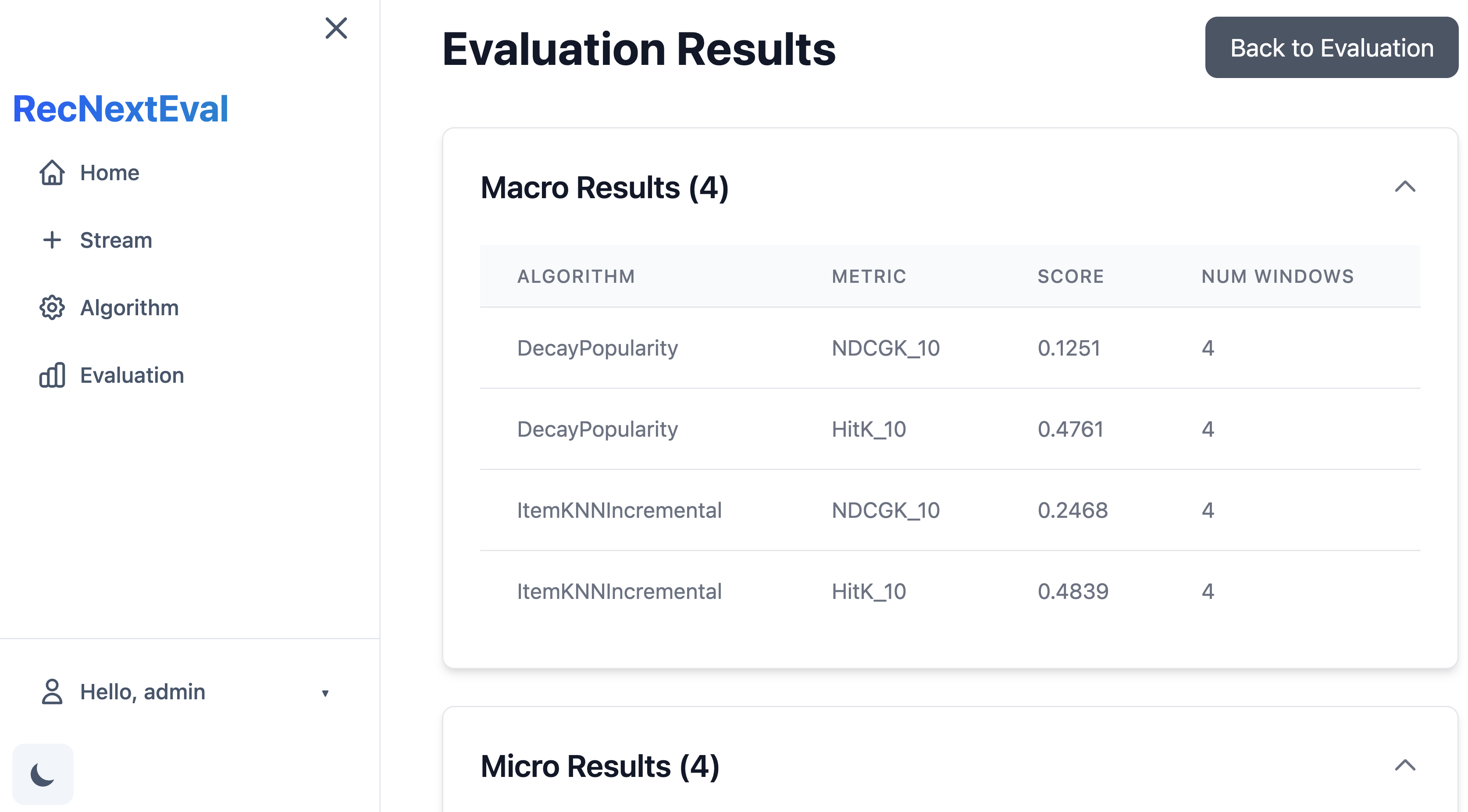}
  \caption{Evaluation metrics reported as macro and micro averages.}
  \Description{Evaluation metrics reported as macro and micro averages.}
  \label{fig:GUI-results}
\end{figure}

We use three sample models (ItemKNNIncremental, RecentPopularity, and DecayPopularity) from our Algorithm package  to demonstrate the capabilities of \modelname. The evaluation is conducted on the MovieLens-100K dataset using two commonly adopted metrics, HitRate@$k$ and NDCG@$k$. We enable the evaluation of both unknown users and unknown items, with an initial temporal split at $t=875156710$ (25 September 1997, 03:05:10). Figures~\ref{fig:GUI-config} and~\ref{fig:GUI-results} show the configuration web interface and the recommendation results measured by NDCG and Hit Rate using macro-averaged metrics.\footnote{Micro-averaged results are omitted from this figure for brevity.}

\begin{figure}
 \center
  \includegraphics[clip, trim=0.3cm 0.3cm 0.3cm 0.3cm, width=0.75\linewidth]{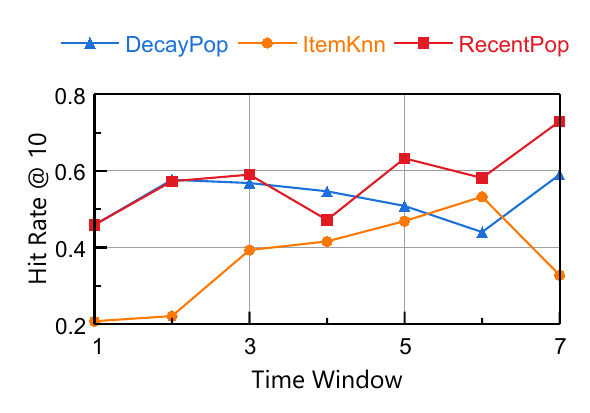}
  \caption{Hit Rate@10 of the three models in 7 time window.}
  \Description{Hit Rate@10 of the three models in 7 time window.}
  \label{fig:WindowResults}
\end{figure}

Figure~\ref{fig:WindowResults} plots the Hit Rate@10 of the three models across seven time windows. As expected, different  algorithms may exhibit varying performance across different windows. Interestingly,  popularity-based algorithms on MovieLens remain competitive when temporal factors are taken into account~\cite{PopularitySigir20}.

\section{Conclusion}
\label{sec:conclude}

We demonstrate \modelname, a reference implementation that focuses exclusively on the evaluation of RecSys models following timeline. 
It is not our objective to develop implementations of specific RecSys algorithms, as adapting each model to incremental learning is non-trivial. Further, both incremental learning and time-windowed evaluation introduce significant computational costs, which require further research.

We argue that \modelname enables a more comprehensive evaluation by considering multiple test instances over time, thereby capturing evolving and dynamic user–item interaction patterns. Moreover, models are given opportunities to improve themselves by incrementally learning from the ground truth revealed after each evaluation phase, as well as from additional training data released over time. We believe that \modelname provides a more concrete platform for RecSys researchers to better understand and assess the real-world behavior of recommender system models.

\balance
\bibliographystyle{ACM-Reference-Format}
\bibliography{RecNextEvaBib}

\end{document}